\documentstyle[prl,floats,aps]{revtex}
\begin{document}
\draft
\twocolumn
\begin{title}
{Strings  in  charge-transfer Mott insulators: effects of lattice
  vibrations and the Coulomb interaction }
\end{title}
\author{A.S. Alexandrov* and V.V. Kabanov**}
\address
{*Department of Physics, Loughborough University, Loughborough LE11
3TU, United Kingdom, **Josef Stefan Institute 1001, Ljubljana, Slovenia}

\maketitle
\begin{abstract}
Applying the canonical transformation with the  $1/\lambda$ perturbation
expansion in the nonadiabatic and intermediate regime and the discrete
generalisation of Pekar's continuous nonlinear  equation in the
extreme adiabatic regime we show that there are no strings in narrow-band ionic
insulators due to the Fr\"ohlich electron-phonon
interaction alone. The multi-polaron system is a homogeneous state
in a wide range of
physically interesting parameters, no matter how strong correlations
are. At the same time the Fr\"ohlich interaction allows  the
antiferromagnetic interactions and/or a short-range electron-phonon
interactions to form short  strings
 in  doped antiferromagnetic insulators if the static dielectric
 constant is large enough.
\end{abstract}

\narrowtext

The electron-phonon interaction is strong in ionic cuprates and manganites
as established both experimentally \cite{mul,tim,mul3,mul2} and theoretically
\cite{alemot,bis,mil,alebra}. The carriers, doped into the  Mott insulator,  are
coupled with the antiferromagnetic background as well. The
antiferromagnetic interactions are  thought to give rise  to  spin
and charge segregation (stripes) \cite{zaa,emkiv}. There is
growing experimental evidence \cite{tran,bia,emkivtran} that stripes occur in
slightly doped  insulators. Their theoretical studies were
restricted so far  to the
repulsive  strongly correlated models \cite{zaa,emkiv},  or  to an extreme
adiabatic limit of the electron-phonon interaction in  narrow
\cite{feo,feo2} and wide band \cite{gri1,gri2} polar semiconductors
and polymers. On the other hand
there is strong evidence that the nonadiabatic electron-phonon
interaction and small
polarons are involed in the  physics of stripes
\cite{mul3,bia}. Also the role of the long-range Coulomb and Fr\"ohlich interactions
remains to  be properly addressed.

In this letter we prove that the Fr\"ohlich electron-phonon
interaction combined with the direct Coulomb repulsion  does not lead
to  charge  segregation like  strings in doped
narrow-band insulators, both in the nonadiabatic and adiabatic regimes.
 However, this
interaction significantly reduces  the Coulomb repulsion, which might allow
 much weaker antiferromagnetic and/or short-range electron-phonon
 interactions to  segregate charges in the doped
  insulators, as suggested by previous studies \cite{zaa,emkiv,feo}.

To begin with, we consider a generic Hamiltonian, including,
respectively,  the
kinetic energy of carriers, the Fr\"ohlich electron-phonon interaction,
phonon energy, and the  Coulomb repulsion as
\begin{eqnarray}
 H &=& \sum_{i \neq j}t({\bf m-n})\delta_{s,s'}
 c^{\dagger}_{i}c_{j}+\sum_{{\bf q},
 i}\omega_{\bf q}n_{i}\left[u_{i}({\bf q})d_{\bf q} + H.c.\right] \cr
 &+&
 \sum_{\bf q}\omega_{\bf q}(d_{\bf q}^{\dagger}d_{\bf q}+1/2)+
 {1\over{2}}\sum_{i\neq j}V({\bf m-n})n_{i}n_{j}
 \end{eqnarray}
with  bare hopping integral $t({\bf m })$, and
 matrix element of the electron-phonon interaction
 \begin{equation}
 u_{i}({\bf q})={1\over{\sqrt{2N}}}\gamma({\bf q})e^{i{\bf q\cdot
 m}}.
 \end{equation}
 Here $i=({\bf m},s)$, $j=({\bf n},s')$ include  site ${\bf m,n}$
 and spin ${s,s'}$ quantum numbers,
$n_{i}=c^{\dagger}_{i}c_{i}$,  $c_{i},d_{\bf q}$ are the
electron (hole)
and phonon operators, respectively, and $N$ is the number of sites.  At large distances ( or small $q$)
 one finds
\begin{equation}
\gamma({\bf q})^2\omega_{\bf q}={4\pi  e^2\over{\kappa q^2}},
\end{equation}
 and
\begin{equation}
V({\bf m-n})={e^2\over{\epsilon_{\infty} |{\bf m-n}|}}.
\end{equation}
 The phonon frequency $\omega_{\bf
 q}$ and the static and high-frequency dielectric constants in $\kappa^{-1}=
 \epsilon_{\infty}^{-1}-\epsilon_{0}^{-1}$ are those of the host insulator
 ($\hbar=c=1$).

   One can apply
 the  canonical transformation \cite{lan} and
  the  $1/\lambda$ multi-polaron perturbation theory \cite{alemot} to integrate out
 phonons,
\begin{equation}
S=\sum_{{\bf q},
 i}n_{i}\left[u_{i}({\bf q})d_{\bf q} -H.c.\right].
 \end{equation}
The result is \cite{alemot,lan}
 \begin{eqnarray}
 \tilde{H}&=&e^SHe^{-S}=\sum_{i\neq j}\hat{\sigma}_{ij}c^{\dagger}_{i}c_{j}
 -E_{p}\sum_{i}n_{i}\cr
&+&\sum_{\bf
 q}\omega_{\bf q}(d_{\bf q}^{\dagger}d_{\bf q}
 +1/2)+{1\over{2}}\sum_{i\neq j}v_{ij}n_{i}n_{j},
 \end{eqnarray}
where
\begin{equation}
\hat{\sigma}_{ij}=t({\bf m-n})\delta_{s,s'}\exp\left(\sum_{\bf q}[u_{i}({\bf q})-u_{j}({\bf q})]d_{\bf q} -H.c.\right)
 \end{equation}
 is the renormalised hopping integral depending on the phonon variables,
 $E_p=zt \lambda$ is the polaron level
 shift and
 \begin{equation}
 v_{ij}=V({\bf m-n}) -{1\over{N}}\sum_{\bf q}\gamma({\bf q})^{2}\omega_{\bf
 q}\cos[{\bf q \cdot (m-n})]
 \end{equation}
is the  net  interaction of polarons
comprising the long-range Coulomb repulsion  and the long-range
attraction due to   ionic lattice deformations.  Here $\lambda= \sum_{\bf q} \gamma({\bf q})^2 \omega_{\bf q}/2Nzt$ is
  the dimensionless coupling constant, $t$ is the nearest neighbour
 hopping integral and $z$ is
  the coordination lattice number.

The extention of the deformation
 surrounding (Fr\"ohlich) polarons  is  large, so
  their
 deformation fields  overlap at finite density. However,   taking
 into account both the long-range attraction of polarons
 due to the  lattice deformations $and$
  the direct Coulomb repulsion, the net
long-range interaction is  repulsive \cite{alemot}.    At
distances larger than the lattice constant, $|{\bf m-n}| \geq a \equiv 1$,  this interaction is significantly reduced  to
\begin{equation}
v_{ij}={e^{2}\over{\epsilon_{0} |{\bf m-n}|}}.
\end{equation}
Optical phonons nearly nullify the bare Coulomb repulsion in ionic
solids if $\epsilon_{0}>>1$, which is normally the case in oxides.
The kinetic energy term in the exact Hamiltonian, Eq.(6) involves
multiphonon events generating a residual $polaron$-phonon interaction
\cite{alemot}.
 Below we show that  in the two opposite limits, the nonadiabatic ($\omega_{\bf q} \geq t$)
and in the extreme adiabatic ( $\omega_{\bf q} \rightarrow 0$) regimes, there is no
charge segregation or any other instability of the polaronic liquid
due to the Fr\"ohlich interaction  in
 doped insulators, but only
  Wigner crystallization at very low
densities.

First we consider the nonadiabtic and intermediate regime. The
properties of a single small polaron with the Fr\"ohlich
electron-phonon interaction were discussed  a long time ago \cite{yam,eag}. Exact
Quantum Monte-Carlo simulations \cite{alekor} showed that the first
order $1/\lambda$ perturbation theory is numerically accurate for
$any$ coupling if the phonon frequency is sufficiently large,
$\omega_{\bf q}>t/2$. The characteristic  frequency
of phonons strongly coupled with carriers is about $\omega_{\bf q}=75$
meV \cite{tim} in cuprates, so cuprates are in this regime.  Hence, one can replace the hopping operator in
Eq.(6) for its phonon average, reducing the problem to narrow-band
fermions with the weak repulsive interaction, Eq.(9).  Next order
corrections in $1/\lambda$ increase
   the polaron binding energy  with  little effect on the
   bandwidth \cite{fir}.
 Because the net long-range repulsion is
 relatively weak, the relevant dimensionless parameter
 $r_{s}=m^{*}e^{2}/\epsilon_{0}(4\pi n/3)^{1/3}$ is not very large in
 doped cuprates.  Wigner
 crystallization appears around $r_{s}\simeq 100$ or larger, which
 corresponds to the atomic density of polarons $n \leq 10^{-6}$ with
  $\epsilon_{0}=30$ and the polaronic mass $m^{*}=5m_{e}$ typical for cuprates and
  manganites.  This estimate  shows
   that small polarons in cuprates and manganites are
      in the homogeneous  state at  physically
   interesting densities.

In the opposite adiabatic limit one can apply a discrete
version  of the continuos nonlinear  equation \cite{pek}
proposed in Ref. \cite{kab} for the Holstein (molecular) model of the electron-phonon
interaction and  extended to the case of the deformation and
Fr\"ohlich interactions in Ref. \cite{feo,feo2}.
Applying the Hartree approximation for the Coulomb repulsion,  the
single-particle  wave-function,
$\psi_{\bf n}$ (the amplitude of the Wannier state $|{\bf n}\rangle$) obeys the following equation
\begin{equation}
-\sum_{{\bf m}\neq 0}t({\bf m})[ \psi_{\bf n}-\psi_{\bf n+m}]-e\phi_{\bf n}
 \psi_{\bf n} = E\psi_{\bf n}.
\end{equation}

The potential $\phi_{{\bf n},k}$ acting on a fermion $k$ at the  site
${\bf n}$  is  created by the
polarization of the lattice $\phi_{{\bf n},k}^{l}$ and by the Coulomb
repulsion with the other $M-1$ fermions,
$\phi_{{\bf n},k}^{c}$,
\begin{equation}
\phi_{{\bf n},k} = \phi_{{\bf n},k}^{l}+\phi_{{\bf n},k}^{c}.
\end{equation}
Both potentials satisfy the descrete Poisson equation as
\begin{equation}
\kappa \Delta \phi_{{\bf n},k}^{l} = 4\pi e
\sum_{p=1}^{M}|\psi_{{\bf n},p}|^{2},
\end{equation}
and
\begin{equation}
\epsilon_{\infty}\Delta \phi_{{\bf n},k}^{c} = -4\pi e
\sum_{p=1,p \neq k}^{M}|\psi_{{\bf n},p}|^{2},
\end{equation}
with $\Delta \phi_{\bf n}=\sum_{\bf m}(\phi_{\bf n}-\phi_{\bf n+m})$.
Differently from Ref. \cite{feo2} we include the Coulomb interaction in
Pekar's functional $J$ \cite{pek}, describing the total energy, in a selfconsistent manner using the Hartree
approximation, so that
\begin{eqnarray}
J&=&-\sum _{{\bf n},p,{\bf m}\neq 0}\psi^*_{{\bf n},p}t({\bf m})
[ \psi_{{\bf n},p}-\psi_{{\bf n+m},p}]\cr
&-&{2\pi e^{2}\over{\kappa}}\sum_{{\bf n},p,{\bf m},q}|\psi_{{\bf n},p}|^{2}\Delta^{-1}
|\psi_{{\bf m},q}|^{2}\cr
& +& {2\pi
  e^{2}\over{\epsilon_{\infty}}}\sum_{{\bf n},p,{\bf m},q \neq p}|\psi_{{\bf n},p}|^{2}\Delta^{-1}
|\psi_{{\bf m},q}|^{2}.
\end{eqnarray}

If we assume, following Ref. \cite{feo} that the single-particle
function of a fermion trapped in a string  of the length $N$ is a simple exponent,
$\psi_{n}=N^{-1/2} \exp (ikn)$
with the periodic boundary conditions, then the functional $J$ is
expressed as $J=T+U$, where $T=-2t(N-1)\sin(\pi M/N)/[N\sin(\pi/N)]$
is the kinetic energy (for an $odd$ number $M$ of spinless fermions) \cite{ref2},  proportional to
$t$, and
\begin{equation}
U=-{e^{2}\over{\kappa}} M^{2} I_{N} + {e^{2}\over{\epsilon_{\infty}}} M(M-1) I_{N},
\end{equation}
corresponds to the polarisation and the Coulomb energies.  Here  the integral $I_{N}$ is given by
\begin{eqnarray}
 I_N&=& {\pi \over{(2\pi)^3}} \int _{-\pi}^\pi dx \int _{-\pi}^\pi dy  \int
     _{-\pi}^\pi dz {\sin(Nx/2)^2\over{N^2 \sin(x/2)^2}}\cr
&\times&( 3-\cos x -\cos
     y -\cos z)^{-1}.
\end{eqnarray}
It has the following asymptotics, Fig.1,
\begin{equation}
I_{N}= {1.31+\ln N\over{N}},
\end{equation}
 which is also derived analyticially at large $N$ by the use of the
 fact that
  $ \sin(Nx/2)^2/(2\pi N\sin(x/2)^2)$ can be replaced by a $\delta$- function.
 If we split the first (attractive) term in Eq.(15) into two parts by
replacing $M^2$ for $M+M(M-1)$, then it  becomes clear that the net interaction between polarons remains repulsive
 in the adiabatic regime as well  because $\kappa
 >\epsilon_{\infty}$. Hence, there are no strings
 within the Hartree approximation for the Coulomb
 interaction. Strong
 correlations  do not change this conclusion. Indeed,
 if we take  the Coulomb energy of spinless one-dimensional fermions comprising both
 Hartree and exchange terms as \cite{ref}
\begin{equation}
  E_C={e^2M(M-1)\over{N\epsilon_{\infty}}} [0.916+\ln M],
\end{equation}
 the polarisation and Coulomb  energy  per particle  becomes (for large $M>>1$)
\begin{equation}
U/M={e^2M\over{N \epsilon_{\infty}}}[0.916+\ln M - \alpha(1.31+\ln N)],
\end{equation}
where $\alpha= 1-\epsilon_{\infty}/\epsilon_0 <1$. Minimising this
energy with respect to the length of the string $N$ we find
\begin{equation}
N= M^{1/\alpha} \exp (-0.31+0.916/ \alpha),
\end{equation}
and
\begin{equation}
(U/M)_{min}= -{e^2\over{\kappa}}
M^{1-1/\alpha} \exp(0.31-0.916 /\alpha).
\end{equation}
Hence,  the potential energy per particle increases
 with the number of particles so that the
energy of $M$ well separated polarons is lower than the energy of  polarons trapped
in a string no matter correlated or not.  The opposite conclusion of
Ref. \cite{feo2} originates in an incorrect approximation of the integral
 $I_N \propto N^{0.15}/N$. The correct asymptotic result is
 $I_N = \ln(N)/N$.

One can argue \cite{feoer} that a  finite  kinetic
energy ($t$) can stabilise a string of a finite length. Unfortunately, this is not correct
either. We performed exact (numerical) calculations of the total
energy $E(M,N)$ of $M$ spinless fermions in a string of the length $N$
including both  kinetic  and potential energy  with the typical
values of $\epsilon_{\infty}=5$ and $\epsilon_{0}=30$. The  local energy minima (per particle)
in the string of the length $1 \leq N \leq 69$ containing $M \leq N/2$
particles are presented in the Table. Strings with the  even fermion numbers carry a finite
current and hence the local minima are found for odd $M$.
   In the  extreme wide
band regime with $t$ as large as 1 eV  the global string  energy  minimum is
found at $M=3, N=25$ ($E= -2.1167$ eV),  and at $M=3, N=13$ for $t=0.5$
eV ($E= -1.2138$ eV). However, this is $not$ the ground state
energy in
both cases. The energy of well separated  $d\geq 2$-dimensional polarons
is well below, less than $-2dt$ per particle (i.e.  $-6$ eV in the first case and $-3$ eV
in the second one in the three dimensional cubic lattice, and $-4$ eV
and $-2$ eV, respectively, in the two-dimensional square lattice). This  argument is applied for any values of $\epsilon_{0},
\epsilon_{\infty}$ and $t$.  As a result we have proved that  strings are
impossible with the Fr\"ohlich interaction alone contrary to the
erroneous Ref. \cite{feo2,feoer}.

The Fr\"ohlich interaction is, of course, not the only electron-phonon
interaction in  ionic solids. As
discussed in Ref. \cite{alemot}, any short range electron-phonon
interaction, like, for example, the  Jahn-Teller (JT) distortion  can overcome
 the residual weak repulsion
of Fr\"ohlich  polarons to form small
bipolarons. At large distances small nonadibatic bipolarons weakly  repel
each other due to the long-range Coulomb interaction, four times of
that of polarons, Eq.(9).   Hence, they  form a liquid state
 \cite{alemot}, or
bipolaronic-polaronic crystal-like structures \cite{aub} depending on their
effective mass and density. The fact, that the Fr\"ohlich interaction
almost nullifies the Coulomb repulsion in oxides justifies the use of
the Holstein-Hubbard model \cite{bis,feh}.   The ground state of the 1D Holstein-Hubbard model is a
liquid of intersite bipolarons with a significantly reduced mass
(compared with the on-site bipolaron) as  shown recently
\cite{bon}.
The  bound states of three or more polarons are not stable in this model,
thus ruling out phase separation. However, the situation might be different
if the antiferromagnetic \cite{zaa,emkiv} and JT interaction\cite{gor} or
 any short (but finite)-range
 electron-phonon interaction   are strong enough.  Due to  long-range
 nature of the Coulomb repulsion the length of a string  should
 be finite (see, also Ref.\cite{bia,feo}). One can readely estimate its length by the
 use of Eq.(8) for any type of the short-range  electron-phonon
 interaction. If, for example, we take dispersive phonons,
 $\omega_{\bf q} = \omega_{0} + \delta \omega (\cos q_{x} + \cos q_{y} +
\cos q_{z})$ with a
 $q$-independent matrix element $\gamma({\bf q})=\gamma$,
we obtain a short-range polaron-polaron attraction as
\begin{equation}
v_{att}({\bf n-m}) = - E_{att} (\delta \omega/\omega) \delta_{|{\bf n-m}|,1},
\end{equation}
where $E_{att}= \gamma^2 \omega_0/2$.
Taking into account the long-range repulsion as well, Eq.(9), the potential energy
of the  string with $M=N$ polarons  becomes
\begin{equation}
U={e^{2}\over{\epsilon_{0}}} N^{2} I_{N} - \frac { N E_{att} \delta \omega} { \omega}.
\end{equation}
Minimization of this energy
yields the length of the string as
\begin{equation}
N = \exp \left(\frac {\epsilon_{0}E_{att} \delta \omega }{e^{2}\omega}
-2.31 \right ).
\end{equation}
Actually, this expression provides a fair estimate of the string
 length   for any kind of attraction (not only generated by
 phonon dispersion),  but also for the  antiferromagnetic  exchange and/or Jahn-Teller type of
 interactions \cite{ref3}.  Due to the numerical coefficient
 in the exponent in Eq.(24) one can expect only  short strings (if
 any) with the realistic values of $E_{att}$ (about a few  hundreds
 millivolts), and the static dielectric constant $\epsilon_{0} \leq 100$.

 We conclude that there are no strings
 in ionic  doped insulators with the  Fr\"ohlich
 interaction alone. Depending on their density and mass polarons remain in a liquid state or  Wigner
 crystal.
On the other hand the short-range electron-phonon  and/or  antiferromagnetic
 interactions might  provide a liquid bipolaronic state and/or   charge segregation (strings of a
 finite length) because the long-range Fr\"ohlich   interaction significantly
 reduces the Coulomb repulsion in highly polarizable ionic insulators.

We  greatly appreciate enlightening  discussions with Antonio
Bianconi, Janez Bonca, Alex Bratkovsky, David Eagles, Pavel
Kornilovitch, Fedor Kusmartsev, Dragan Mihailovic, and Jan Zaanen.
One of us (V.V.K.) acknowledges support of the work by INTAS grant
No.97-963.

\centerline{\bf Figure captions}

Fig.1. The polarisation energy of small Fr\"ohlich polarons trapped in a string
depends on its length as $\ln(N)/N$.
\begin{table}[tbp]
\caption{$E(M,N)$ for $t=1$ eV and $t=0.5$ eV}
\begin{tabular}[t]{lllll}
 & $t=1$eV &  & $t=0.5$eV & \\ \hline
M & N  & E(M,N)  & N & E(M,N) \\ \hline
1 & 11 & -2.0328 & 3 & -1.1919 \\
3 & 25 & -2.1167 & 13 & -1.2138 \\
5 & 42 & -2.1166 & 25 & -1.1840 \\
7 & 61 & -2.1127 & 40 & -1.1661 \\
\end{tabular}
\end{table}

\begin{references}
\bibitem{mul}
G. Zhao $et$ $al$,  Nature ${\bf
 385}$, 236 (1997).
\bibitem{tim}
T.\,Timusk {\em et al}, in
{\em Anharmonic Properties of High-$T_{c}$ Cuprates},
eds. D.\,Mihailovi\'c {\em et al},
(World Scientific, Singapore, 1995), p.171.
\bibitem{mul3}
A. Lanzara, $et$ $al$, Journ. Phys.: Condens. Mat. {\bf 11}, L541 (1999)
\bibitem{mul2}
D.R. Temprano $et$ $al$, Phys. Rev. Lett. {\bf 84}, 1982 (2000).
\bibitem{alemot}
A.S.  Alexandrov and N.F. Mott, Rep.Prog.Phys. ${\bf 57}$, 1197
(1994;
 {\em `Polarons and Bipolarons'}, World
Scientific (1995).
\bibitem{bis}
A.R. Bishop  and M. Salkola , in:
 `Polarons and Bipolarons in High-$T_{c}$ Superconductors and
 Related Materials', eds E.K.H. Salje, A.S. Alexandrov  and W.Y.  Liang,
Cambridge University Press, Cambridge, 353 (1995);
\bibitem{mil}
A.J. Millis {\em et al.}, Phys. Rev. Lett. {\bf 74}, 5144 (1995);
\bibitem{alebra}
A.S. Alexandrov and A.M. Bratkovsky, Phys. Rev. Lett. {\bf 82}, 141
(1999); ibid {\bf 84}, 2043 (2000).
\bibitem{zaa}
J. Zaanen and O. Gunnarsson, Phys. Rev. B {\bf 40}, 7391 (1989).
\bibitem{emkiv}
V. J. Emery $et$ $al$, Phys. Rev. B {\bf 56}, 6120 (1997) and
refrences therein.
\bibitem{tran}
J.M. Tranquada $et$ $al$, Nature {\bf 375}, 561 (1996).
\bibitem {bia}
A. Bianconi, J. Phys. IV France, {\bf 9}, 325 (1999) and references therein.
\bibitem{emkivtran}
V. J. Emery $et$ $al$, to appear in {\it Proc. Natl. Acad. Sci USA},
preprint cond-mat/9907228.
\bibitem{feo}
F.V. Kusmartsev, J. Phys. IV France, {\bf 9}, 321 (1999).
\bibitem{feo2}
F.V. Kusmartsev, Phys. Rev. Lett., {\bf 84}, 530 (2000).
\bibitem{gri1}
L.N. Grigorov, Makromol. Chem., Macromol. Symp. {\bf 37}, 159 (1990).
\bibitem{gri2}
L.N. Grigorov $et$ $al$,  Makromol. Chem., Macromol. Symp. {\bf 37},
177 (1990).
\bibitem{lan}
I.G.  Lang  and Yu.A. Firsov, Zh. Eksp. Teor. Fiz. ${\bf 43}$, 1843 (1962)
( Sov. Phys. JETP ${\bf 16}$, 1301 (1963)).
\bibitem{yam}
J. Yamashita and T. Kurosawa, J. Phys. Chem. Solids ${\bf 5}$, 34 (1958).
\bibitem{eag}
D.M. Eagles, Phys. Rev. {\bf 181}, 1278 (1969) and references therein.
\bibitem{alekor}
A.S. Alexandrov and P.E. Kornilovich,  Phys. Rev. Lett. ${\bf 82}$,
807 (1999).
\bibitem{fir}
Yu.A. Firsov $et$ $al$, Phys. Rev. B ${\bf 59}$, 12132 (1999).
\bibitem{pek}
S.I. Pekar, Zh. Eksp. Teor. Fiz. ${\bf 16}$, 335 (1946).
\bibitem{kab}
V.V. Kabanov and O.Yu. Mashtakov, Phys. Rev. B${\bf 47}$, 6060 (1993).
\bibitem{ref}
This expression differs from
  Ref. \cite{feo,feo2} by  the  numerical coefficients.
\bibitem{ref2}
For an $even$ $M$ the kinetic energy is $T=-2t(N-1)\sin(\pi
M/N)\cos(\pi/N)/[N\sin(\pi/N)]$.
\bibitem{feoer}
F.V. Kusmartsev, Phys. Rev. Lett. Erratum, {\bf 84}, 5026 (2000).
\bibitem{aub}
S. Aubry, in:
 `Polarons and Bipolarons in High-$T_{c}$ Superconductors and
 Related Materials', eds E.K.H. Salje, A.S. Alexandrov  and W.Y.  Liang,
Cambridge University Press, Cambridge, 271 (1995).
\bibitem{feh}
H. Fehske $et$ $al$,  Phys. Rev.
 B{\bf 51}, 16582 (1995).
\bibitem{bon}
J. Bonca $et$ $al$, Phys. Rev. Lett., {\bf 84}, (2000).
\bibitem{gor}
L.P. Gor'kov and A.V. Sokol, Pis'ma Zh. Eksp. Teor. Fiz., ${\bf 46}$,
333 (1987).
\bibitem{ref3}
$E_{att}\delta \omega/\omega  \propto J$ in the case of the 
antiferomagnetic interaction
$J$, and $E_{att}\delta \omega/\omega \propto E_{JT}$ in the case of the 
JT distortion ($E_{JT}$
is the Jahn-Teller energy).


\end{references}
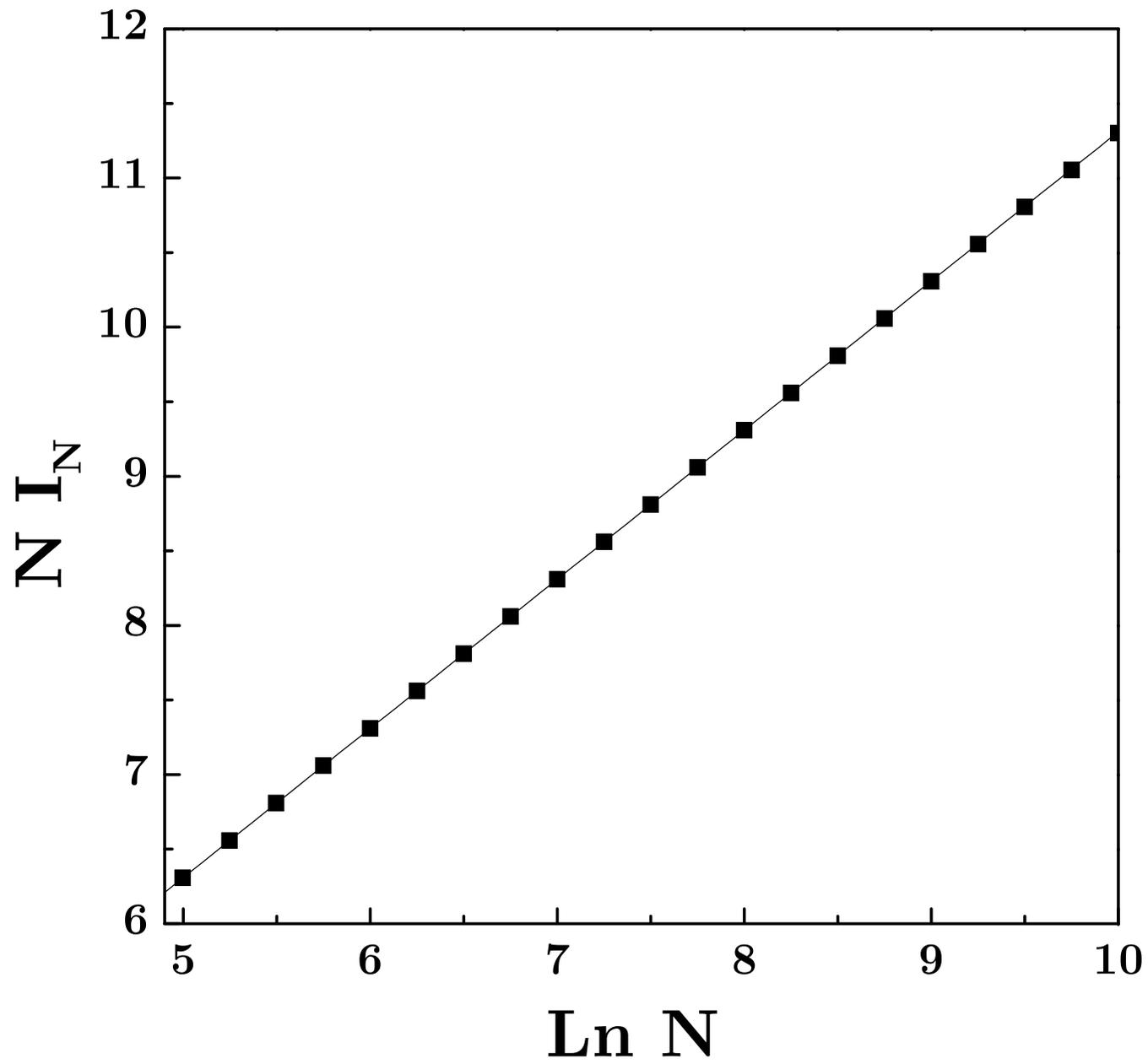
\end{document}